\documentclass{article}%
\usepackage{amssymb}
\usepackage{graphicx}
\usepackage{amsmath}
\usepackage{amsfonts}%
\providecommand{\U}[1]{\protect\rule{.1in}{.1in}}

\begin{document}

\begin{center}
Solution of the\ asymmetric double sine-Gordon equation

Nan-Hong Kuo$^{1}$, Sujit Sarkar$^{2}$, C. D. Hu$^{1,3}$

$^{1}$Department of physics, National Taiwan University, Taipei, Taiwan, R.O.C.

$^{2}$Poornaprajna Institute of Scientific Research, 4 Sadashivanagar,
Bangalore-560 080, India.

$^{3}$Center for Theoretical Sciences, National Taiwan University, Taipei,
Taiwan, R.O.C.

Abstract
\end{center}

We present solutions of asymmetric double sine-Gordon equation (DSGE) of an
infinite system based on M\"{o}bius transformation and numerical exercise.
This method is able to give the forms of the solutions for all the region on
the $\varphi-\eta$ parameter plane where $\varphi$\ is an additional phase and
$\eta$\ is the ratio of the magnitudes of two sine terms.\ We are able to show
how the deconfinement occurs near\ $\varphi=(1/2+n)\pi$\ and $\varphi=n\pi.$
and also find the solution for all values of $\varphi.$\ We predict different
kind of solutions and transitions among them in different parts of the
parameter space of this equation.\newpage

\subsubsection{1. Introduction:}

The sine-Gordon equation attracted much interest of physicists[1,2]. In
quantum field theory it is a rare example of integrable system which can be a
starting point of developing non-perturbative theory. It also has plenty of
application in condensed matter systems[3,4] and nonlinear optics[5]. However,
the Lagrangian of a realistic physical system often gives a more complicated
equation of motion than the sine-Gordon equation. For example, a quantum spin
chain is mapped into a Lagrangian with several potential terms[6]. Systems
with nonlinear optical properties also give rise to more complicated wave
equations[7]. Thus a more complete model is desirable. This leads to the
double sine-Gordon equation (DSGE) and loses the integrability of the
sine-Gordon equation. But this "flaw" also provides a test ground for various
analysis[8-11] and perturbative or non-perturbative methods such as form
factor perturbation theory[12,13],.truncated conformal space approach[13,14]
and semiclassical approach[15]. For a standard DSGE, exact solutions can be
found[8,10]. But for a asymmetric DSGE%
\begin{equation}
\theta_{tt}-\theta_{xx}+\sin(\theta+\varphi)+2\eta\sin2\theta=0. \tag{1}%
\end{equation}
with $\varphi$ being an additional phase, exact solution has eluded the effort
of researchers. It is interesting not only because the state equations of a
strongly correlated electron system[16], a quantum spin chain with external
field or that of a spin pump[17,18] can be mapped into eq. (1) but from a
purely theoretical point of view, it contain rich physics as we shall see.

The main difficulty of DSGE is that it is non-integrable. A rigorous
analytical form of the solutions of asymmetric DSGE is important because it
can provide deeper insight to understand the physical systems. It is also well
known that the perturbation theory is not applicable in one-dimension because
there is no quasi-particle excitation alike in higher dimension. Various
approximation methods, perturbative or non-perturbative, sometimes give
results leading to different physical pictures[14]. For this reasons, we would
not like to solve this equation with above methods. We propose an insightful
analytical form of the solution which can serve as an anchor to numerical
analysis or approximations. It can also be a springboard to study quantum fluctuations.

The plan of the manuscript is the following: In Section \textbf{2}, we present
previously existed classical solutions of DSGE depend on the potential
coefficient $\eta$. We also present explicitly shortcomings of these
analytical solutions. Section \textbf{3} contains the major part of our work.
We present mathematical solution we find with M\"{o}bius transformation and
see how parameters change with $\eta$ and $\varphi.$ In Section \textbf{4}, We
present results and discussions.

\subsubsection{2. Classical solutions of Double Sine-Gordon Equation}

In this section, we briefly review the previously existed solutions in the
literature, potential and energy of the double sine-Gordon equation (DSGE)
with $\varphi=0$ in order to establish the notations and set the stage for
developing our method. We shall use the notations of ref. 4 and also point out
the shortcoming of that work. As mentioned in ref. 4, the Hamiltonian of the
DSGE can be viewed as a chain of physical pendulums joined by torsion springs
\begin{equation}
H=%
{\displaystyle\int}
dx\{\frac{p_{\theta}^{2}}{2I}+\frac{\Gamma}{2}\theta_{x}^{2}-V_{0}[\cos
\theta+\eta\cos2\theta]\} \tag{2}%
\end{equation}
where $\theta$\ is the angular coordinate, $p_{\theta}$\ is the conjugate
momentum, $I$\ is the moment of inertia, $\Gamma$\ is the torsional constant,
and $V_{0}$ is the external potential. The DSGE which we are interested in
(eq. (1).) can be obtained with rescalings of the time-space coordinates. In
order to calculate the total energy, one can Integrate both sides of eq. (1)
and get
\begin{equation}
\frac{1}{2}(\frac{d\theta}{ds})^{2}+\cos(\theta+\varphi)+\eta\cos(2\theta)=S
\tag{3}%
\end{equation}
where $s=\gamma(x-vt)$ , $\gamma=1/\sqrt{1-v^{2}}$ and $\nu$ is the velocity
of soliton. The negative sum of the last two terms can be viewed as the
potential
\begin{equation}
V(\theta;\eta,\varphi)=-\cos(\theta+\varphi)-\eta\cos(2\theta) \tag{4}%
\end{equation}
and so $S$ is called \ the \textquotedblright action\textquotedblright. We can
show that
\begin{equation}
S=-V_{\min}. \tag{5}%
\end{equation}
The solutions of kinks and bubbles (see ref. 4 and/or eqs. (28), (35) and (37)
below) approach constant when $s\rightarrow\pm\infty$. Hence, $d\theta
/ds|_{s=\pm\infty}=0$ and $S=-V(\theta(\pm\infty))$. Since $(d\theta
/ds)^{2}/2=V(\theta)-V(\theta(\pm\infty))\geq0$\ for any $s$, we conclude that
$S=-V(\theta(\pm\infty))=-V_{\min}$. More specifically, whether the minimum is
an absolute minimum or a relative minimum depends on what type of the solution is.

Now we would like to calculate energy for all $\varphi$. The calculation of
energy can be performed with the a method similar to that of ref. 2. We get
from eq. 3. that%
\begin{equation}
\frac{d\theta}{ds}=\sqrt{2[S-\cos(\theta+\varphi)-\eta\cos2\theta]}. \tag{6}%
\end{equation}
We define
\begin{equation}
V_{s}\equiv\int_{-\infty}^{\infty}[\frac{\theta_{x}^{2}}{2}+\frac{\theta
_{t}^{2}}{2}]dx=(1+v^{2})\gamma\int_{-\infty}^{\infty}\frac{\theta_{s}^{2}}%
{2}ds=\frac{(1+v^{2})\gamma}{\sqrt{2}}\int_{0}^{2\pi}\sqrt{S-\cos
(\theta+\varphi)-\eta\cos2\theta}d\theta\tag{7}%
\end{equation}
and
\begin{equation}
V_{p}\equiv\int_{-\infty}^{\infty}[-\cos(\theta+\varphi)-\eta\cos
2\theta]dx=\frac{1}{\gamma}\int_{-\infty}^{\infty}[-\cos(\theta+\varphi
)-\eta\cos2\theta]ds. \tag{8}%
\end{equation}
so that $H=V_{s}+V_{p}$ with some scaling. From eq. (3) we found that
\begin{equation}
V_{p}=-\frac{S\cdot s|_{s=0}^{s=L}}{\gamma}+\frac{1}{\gamma}\int_{-\infty
}^{\infty}\frac{\theta_{s}^{2}}{2}ds. \tag{9}%
\end{equation}
But the first term on the right hand side of eq. (8) diverges if we consider
an infinite system ($L\rightarrow\infty$). Thus we subtract this trivial
infinity from the Hamiltonian in eq. (2). This is equivalent to shift the
system to a new energy zero. In view of eq. (2), we get%
\begin{equation}
H=V_{s}+V_{p}=2\gamma\int_{-\infty}^{\infty}\frac{\theta_{s}^{2}}{2}ds.
\tag{10}%
\end{equation}
While solving for solutions, we can obtain numerical values of $S$. Hence, the
energy at every $\varphi$ can be calculated.

An analysis of the potential term is in order. It helps us to understand how
state evolves with respect to the phase $\varphi$. The potential of double
sine-Gordon equation has the symmetry%
\begin{equation}
V(\theta;\eta,\varphi=\frac{\pi}{2})=V(\widetilde{\theta};-\eta,\varphi=0)
\tag{11}%
\end{equation}
where $\widetilde{\theta}=\theta+\pi/2.$ This implies that the solution of
$\varphi=\pi/2$\ has the same form as that of $\varphi=0$\ and $-\eta$.
However, interestingly, the solutions can have completely different forms for
$\varphi=0$\ or $\varphi=\pi/2$\ when $|\eta|>\frac{1}{4}$ (see below).

In view of eq. (4),\ we have
\begin{align*}
\frac{dV}{d\theta}  &  =\sin(\theta+\varphi)+2\eta\sin(2\theta)\\
&  =\sin\theta(\cos\varphi+\cot\theta\sin\varphi+4\eta\cos\theta)\\
&  =\cos\theta(\sin\varphi+\tan\theta\cos\varphi+4\eta\sin\theta).
\end{align*}
The point $\theta=\cos^{-1}(-1/4\eta)$ is one of the absolute maxima only if
$\varphi=0$ and $\eta<-1/4$, and $\theta=\sin^{-1}(-1/4\eta)$ is one of the
absolute minima only if $\varphi=\pi/2$ and $\eta>1/4$. Moreover, for
$\eta<-1/4$ the relative minima are\ at $\theta=\pi/2+2n\pi$ and for
$\eta>1/4$\ the relative maxima are at $\theta=2n\pi$. The shape of the
potential has critical influence on the form of the solution. We show their
features in Figures 1a\symbol{126}1c versus $\theta$ for $\eta=-0.35,$
$\eta=0.15$ and $\eta=0.35$ respectively. Notably, for $(\eta=-0.35,\varphi
=\pi/2)$\ and $(\eta=0.35,\varphi=0)$\ there are relative minima and for
$(\eta=-0.35,\varphi=0)$\ and $(\eta=0.35,\varphi=\pi/2)$\ there are relative
maxima.\ Nevertheless, for $\eta=0.15,$\ there is no relative extremum. In
general, there is a region where no relative extremum exists. In the regions
where\ relative extremum exist, a so called bubble solution can be found.
Another symmetry can be seen by making transformation $\theta^{\prime}%
=\theta+\varphi$, eq. (4) becomes $V(\theta;\eta,\varphi)=-\cos\theta^{\prime
}-\eta\cos(2\theta^{\prime}-2\varphi)$. This shows that there is a period of
$\pi$ for the variation of $\varphi$.

Specifically, when $\varphi=0$ and $\varphi=\pi/2$\ the analytical forms of
solution and energy can be obtained. In order to set up our analysis, in the
following we give a summary of\ the solutions and energies in the case
$\varphi=0$.

Case 1 $\eta<-\frac{1}{4}$

In this case the absolute minima are at $\theta_{abs.\min}=\varphi_{0}+2n\pi$
where
\begin{equation}
\varphi_{0}\equiv\arccos(\frac{-1}{4\eta}). \tag{12}%
\end{equation}
From above, we have $S=-V(\theta_{\min})=-\eta-1/8\eta$. The absolute maximum
of $V(\theta)$ are located at $\theta_{abs.\max}=(2n+1)\pi$ while the relative
maximum are located at $\theta_{rel.\max}=2n\pi$\ with $V(\theta_{abs.\max
})=1-\eta$ and $V(\theta_{rel.\max})=-1-\eta.$\ There are two kinds of
traveling kinks:
\begin{equation}
\theta^{>}=2\arctan[\pm\sqrt{\frac{4|\eta|-1}{4|\eta|+1}}\coth(\sqrt
{\frac{16\eta^{2}-1}{16|\eta|}}s)], \tag{13}%
\end{equation}%
\begin{equation}
\theta^{<}=2\arctan[\pm\sqrt{\frac{4|\eta|-1}{4|\eta|+1}}\tanh(\sqrt
{\frac{16\eta^{2}-1}{16|\eta|}}s)] \tag{14}%
\end{equation}
where the superscripts
$>$
and
$<$
denote the large kink and small kink respectively. We discuss their energies separately.

(A) Large kink: We found from eq. (13) that $\theta\in\lbrack\varphi_{0}%
,2\pi-\varphi_{0}]$, and it must vary cross one of the absolute maxima. In
view of eqs. (7) and (10),
\begin{align}
\left\langle H\right\rangle  &  =2\gamma\int_{-\infty}^{\infty}\frac
{\theta_{s}^{2}}{2}ds=\sqrt{2}\gamma\int_{\varphi_{0}}^{2\pi-\varphi_{0}}%
\sqrt{-\eta-\frac{1}{8\eta}-\cos\theta-\eta\cos2\theta}d\theta=\sqrt{2}%
\gamma\int_{\varphi_{0}}^{2\pi-\varphi_{0}}\sqrt{-2\eta(\cos\theta+\frac
{1}{4\eta})^{2}}d\theta\nonumber\\
&  =\frac{\gamma}{\sqrt{-\eta}}(\sqrt{16\eta^{2}-1}+\pi-\varphi_{0}) \tag{15}%
\end{align}

(B) Small kink: From eq. (14) we found that $\theta\in\lbrack-\varphi
_{0},\varphi_{0}]$, and it must contain one of the relative maxima
\begin{equation}
\left\langle H\right\rangle =2\gamma\int_{-\infty}^{\infty}\frac{\theta
_{s}^{2}}{2}ds=\sqrt{2}\gamma\int_{\varphi_{0}}^{-\varphi_{0}}\sqrt
{-\eta-\frac{1}{8\eta}-\cos\theta-\eta\cos2\theta}d\theta=\frac{\gamma}%
{\sqrt{-\eta}}(\sqrt{16\eta^{2}-1}-\varphi_{0}) \tag{16}%
\end{equation}
Here, we have corrected the errors in eqs.(3.10) and (3.11) of ref. 4.

\noindent Case 2\ $|\eta|<\frac{1}{4}$

There is only one type of basic kink solution in this case:
\begin{equation}
\theta^{>}=2\arctan[\pm\sqrt{1+4\eta}csch(\sqrt{1+4\eta}s)] \tag{17}%
\end{equation}
The minimum of $V(\theta)$ are located at $\theta_{\min}=2n\pi$
with$\ V(\theta_{\min})=-1-\eta$ and the maximum of $V(\theta)$ are located at
$\theta_{\max}=(2n+1)\pi$ with\ $V(\theta_{\max})=1-\eta.$ There is no
relative maximum or minimum. We have $S=-V(\theta_{\min})=1+\eta$ and from
eqs. (6) and (9),
\begin{equation}
\left\langle H\right\rangle =2\gamma\int_{-\infty}^{\infty}\frac{\theta
_{s}^{2}}{2}ds=\sqrt{2}\gamma\int_{0}^{2\pi}\sqrt{1+\eta-\cos\theta-\eta
\cos2\theta}d\theta\tag{18}%
\end{equation}
The integral also depends on wether $\eta$ is larger or smaller than $0$.

(A)\textbf{\ }$\frac{1}{4}>\eta>0$%
\begin{equation}
\left\langle H\right\rangle =2\gamma\int_{-\infty}^{\infty}\frac{\theta
_{s}^{2}}{2}ds=4\gamma\sqrt{4\eta+1}+\frac{2\gamma\ln(2\sqrt{\eta}+\sqrt
{4\eta+1})}{\sqrt{\eta}} \tag{19}%
\end{equation}

(B)\textbf{\ }$0>\eta>-\frac{1}{4}$%
\begin{equation}
\left\langle H\right\rangle =2\gamma\int_{-\infty}^{\infty}\frac{\theta
_{s}^{2}}{2}ds=4\gamma\sqrt{4\eta+1}+\frac{2\gamma\arcsin(2\sqrt{-\eta}%
)}{\sqrt{-\eta}} \tag{20}%
\end{equation}
Here we have corrected an error in eq. (3.7) of ref. 4.

Case 3\ $\eta>\frac{1}{4}$

There are two kinds of traveling kinks:
\begin{equation}
\theta^{>}=2\arctan[\pm\sqrt{1+4\eta}csch(\sqrt{1+4\eta}s)], \tag{21}%
\end{equation}
and
\begin{equation}
\theta^{B}=2\arctan[\pm\frac{1}{\sqrt{4\eta-1}}\cosh(\sqrt{4\eta-1}s)]
\tag{22}%
\end{equation}
where the superscript B denotes the bubble solution. The absolute minimum of
$V(\theta)$ are located at $\theta_{abs.\min}=2n\pi$ with $V(\theta_{abs.\min
})=-1-\eta$ and the maximum of $V(\theta)$ are located at $\theta_{\max
}=\arccos(-1/4\eta)+2n\pi$ with $V(\theta_{abs.\max})=1/8\eta+\eta$. The
energy of the large kink is the same as that in case 2:%

\begin{equation}
\left\langle H\right\rangle =2\gamma\int_{-\infty}^{\infty}\frac{\theta
_{s}^{2}}{2}ds=4\gamma\sqrt{4\eta+1}+\frac{2\gamma\ln(2\sqrt{\eta}+\sqrt
{4\eta+1})}{\sqrt{\eta}} \tag{23}%
\end{equation}

The other kind of solution is the bubble solution. It extends from one
relative minimum to another. These minima are at $\theta_{rel\min}%
\rightarrow(2n+1)\pi$ as $s\rightarrow\pm\infty$. In this case, $S=-V(\theta
_{rel,\min})=\eta-1$%
\begin{align}
\left\langle H\right\rangle  &  =2\gamma\int_{-\infty}^{\infty}\frac
{\theta_{s}^{2}}{2}ds=2\sqrt{2}\gamma\int_{2\arctan(1/\sqrt{4\eta-1})}^{\pi
}\sqrt{\eta-1-\cos\theta-\eta\cos2\theta}d\theta\tag{24}\\
&  =4\gamma\sqrt{4\eta-1}-\frac{4\gamma\ln(2\sqrt{\eta}+\sqrt{4\eta-1})}%
{\sqrt{\eta}}.\nonumber
\end{align}
Note the upper bound and lower bound of the integral. We divide the bubble
into two equal halves. The upper bound $\pi$ is the relative minimum
while\ the lower bound is the middle point of the bubble. It comes from
$\theta^{B}(s=0)=2\arctan[\pm(4\eta-1)^{-1/2}\cosh(\sqrt{4\eta-1}%
s)]|_{s=0}=\pm2\arctan(1/\sqrt{4\eta-1})$. Here we also corrected the errors
in eq.(3.7) and (3.9) of ref. 4.

Further insight can be gained by applying eq. (11). For example, if we start
from $\varphi=0$ and $\eta>\frac{1}{4}$, the solutions are the large kink in
eq. (21) and bubble in eq. (22). When $\varphi$ is changed adiabatically into
$\pi/2$, the solutions become the large and small kinks in eqs. (13) and
(14):
\[
\theta^{>}=2\arctan[\pm\sqrt{\frac{4|\eta|-1}{4|\eta|+1}}\coth(\sqrt
{\frac{16\eta^{2}-1}{16|\eta|}}s)]-\frac{\pi}{2},
\]%
\[
\theta^{<}=2\arctan[\pm\sqrt{\frac{4|\eta|-1}{4|\eta|+1}}\tanh(\sqrt
{\frac{16\eta^{2}-1}{16|\eta|}}s)]-\frac{\pi}{2}%
\]
with the additional term $-\pi/2$ coming from the difference between $\theta$
and $\widetilde{\theta}$. Therefore, the solutions can have quite different
forms as $\varphi$\ varies.\ The interesting question is whether the solution
evolve smoothly or they change abruptly.

\subsubsection{3. Solutions in general}

In previous studies of asymmetric DSGE, the solution can be found when
$\varphi=0$\ or $\varphi=\pi/2.$\ Here, we present a method which enables us
to find solutions for any value of $\varphi.$\ We propose that the solution of
eq. (1) in general has the form
\begin{equation}
\theta=2\arctan[f(s)]. \tag{25}%
\end{equation}
Then after substitution, we have the following equation
\begin{equation}
2(\frac{df}{ds})^{2}=(S+\cos(\varphi)-\eta)f^{4}+2\sin(\varphi)f^{3}%
+(2S+6\eta)f^{2}+2\sin(\varphi)f+(E-\cos(\varphi)-\eta) \tag{26}%
\end{equation}
Above equation is similar to the differential equation of Jacobi elliptic
functions (JEF) or hyperbolic functions except for the terms with the odd
power of $f(s).$ However, for JEF, there is the \textquotedblright M\"{o}bius
transformation\textquotedblright\ to change eq. (26) into the standard form.
The details is given in Appendix. By letting $f(s)=(ag(s)+b)/(cg(s)+d)$ and
choosing suitable coefficients, $a$, $b$, $c$, and $d$, it is possible to
obtain the following form from eq. (26)
\begin{equation}
(\frac{dg}{ds})^{2}=a_{4}g^{4}+a_{2}g^{2}+a_{0}. \tag{27}%
\end{equation}

In an infinite system with arbitrary value of $\varphi$, it is reasonable to
use for $g(s)$ the hyperbolic functions which JEFs approach in the $\lim$it of
modulus $k\rightarrow1$:
\begin{equation}
f(s)=\frac{a\sinh(rs)+b}{c\sinh(rs)+d} \tag{28}%
\end{equation}
where $r$\ is a constant. One may also use the other hyperbolic function for
this construction. Substituting eqs. (25) and (28) into eq. (26) and requiring
the same scaling
\begin{equation}
(ad-bc)^{2}=1 \tag{29}%
\end{equation}
we have found the following equations by comparing the powers of $\sinh$
function
\begin{equation}
(a^{2}+c^{2})^{2}S+(a^{4}-c^{4})\cos\varphi+2ac(a^{2}+c^{2})\sin
\varphi+(-a^{4}+6a^{2}c^{2}-c^{4})\eta=0, \tag{30a}%
\end{equation}%
\begin{multline}
4(ab+cd)(a^{2}+c^{2})S+4(a^{3}b-c^{3}d)\cos\varphi\nonumber\\
+2(a^{3}d+3a^{2}bc+3ac^{2}d+bc^{3})\sin\varphi\nonumber\\
+[-4a^{3}b+6(2a^{2}cd+2abc^{2})-4c^{3}d]\eta=0, \tag{30b}%
\end{multline}%
\begin{multline}
\lbrack6a^{2}b^{2}+2(a^{2}d^{2}+4abcd+b^{2}c^{2})+6c^{2}d^{2}]S\nonumber\\
+(6a^{2}b^{2}-6c^{2}d^{2})\cos\varphi\nonumber\\
+6(a^{2}bd+ab^{2}c+acd^{2}+bc^{2}d)\sin\varphi\nonumber\\
+6(-a^{2}b^{2}+a^{2}d^{2}+4abcd+b^{2}c^{2}-c^{2}d^{2})\eta=2r^{2} \tag{30c}%
\end{multline}%
\begin{multline}
4(ab+cd)(b^{2}+d^{2})S+4(b^{3}a-d^{3}c)\cos\varphi+\nonumber\\
2(b^{3}c+3b^{2}ad+3bd^{2}c+ad^{3})\sin\varphi+\nonumber\\
\lbrack-4b^{3}a+6(2d^{2}ab+2cdb^{2})-4d^{3}c]\eta=0 \tag{30d}%
\end{multline}%
\begin{equation}
(b^{2}+d^{2})^{2}S+(b^{4}-d^{4})\cos\varphi+2bd(b^{2}+d^{2})\sin
\varphi+(-b^{4}+6b^{2}d^{2}-d^{4})\eta=2r^{2} \tag{30e}%
\end{equation}
The simultaneous algebraic equations are solved to give the coefficients $a$,
$b$, $c$, $d$, $r$ and $S$. Thus, we can find solutions for any values of
$\eta$\ and\ $\varphi$ except for certain special cases with the form in eq.
(28). In Table 1, 2 and 3 we give the values of the parameters for several
values of $\eta$ and $\varphi$. The crosses in Table 3 indicate that eq. (28)
cannot give any solution under these conditions.

The special case for which the form in eq. (28) cannot give any solution is
well-defined and very interesting, it occurs at $\varphi=\pi/2$ and $\eta
>1/4$. We also explain the cause of this forbidden solution by starting with
the symmetry of the DSGE with additional phase. If one substitutes
$\pi-\varphi$ for $\varphi$, then he can replace $\theta$ with $\pi-\theta
$\ and eq. (1) retains its original form. In view of eqs. (25) and (28),\ we
have the following symmetry:
\begin{equation}%
\begin{array}
[c]{ccc}%
\varphi & \rightarrow & \pi-\varphi\\
a & \rightarrow & -c\\
b & \rightarrow & d\\
c & \rightarrow & -a\\
d & \rightarrow & b\\
r & \rightarrow & r\\
S & \rightarrow & S
\end{array}
. \tag{31}%
\end{equation}
In other words, when $\varphi$\ is changed into $\pi-\varphi,$\ the solution
becomes $\pi-\theta$. Suppose we have, and indeed we have found, real solution
of eqs. (30) for $\eta>1/4$ with $a$,\ $b$,\ $c$\ and $d$\ being continuous
with varying $\varphi$, then we must have $a=-c$ and $b=d$ at $\varphi=\pi/2$.
Substituting these into eqs. (30), we get
\begin{equation}
a^{4}(S-1+\eta)=0, \tag{32a}%
\end{equation}%
\begin{equation}
a^{2}b^{2}=\frac{r^{2}}{4S-12\eta}, \tag{32b}%
\end{equation}%
\begin{equation}
b^{4}(S+1+\eta)=\frac{r^{2}}{2}, \tag{32c}%
\end{equation}
along with the scaling restriction from eq. (29)
\begin{equation}
4a^{2}b^{2}=1. \tag{33}%
\end{equation}
Eq. (33) means $a\neq0$, \ and so eq. (32a)\textbf{\ }implies $S+\eta=1$.
Inserting these into eqs. (32b) and (32c), we get
\begin{equation}
\frac{r^{2}}{1-4\eta}=1. \tag{34}%
\end{equation}
Incompatibility results from\textbf{\ }eq. (32) if $\eta>1/4$.

On the other hand, we can find a solution of distinctively different form at
$\eta>1/4$ and $\varphi=\pi/2$. Instead of $g(s)=\sinh(rs)$, we now have
$g(s)=\tanh(rs)$ or $g(s)=\coth(rs)$%
\begin{equation}
f(s)=\frac{a\sinh(rs)+b\cosh(rs)}{c\sinh(rs)+d\cosh(rs)}. \tag{35}%
\end{equation}
The algebraic equations are
\begin{equation}
(a^{2}+c^{2})^{2}S+(a^{4}-c^{4})\cos\varphi+2ac(a^{2}+c^{2})\sin
\varphi+(-a^{4}+6a^{2}c^{2}-c^{4})\eta=2r^{2} \tag{36a}%
\end{equation}%
\begin{multline}
4(ab+cd)(a^{2}+c^{2})S+4(a^{3}b-c^{3}d)\cos\varphi\nonumber\\
+2(a^{3}d+3a^{2}bc+3ac^{2}d+bc^{3})\sin\varphi\nonumber\\
+[-4a^{3}b+6(2a^{2}cd+2abc^{2})-4c^{3}d]\eta=0 \tag{36b}%
\end{multline}%
\begin{multline}
\lbrack6a^{2}b^{2}+2(a^{2}d^{2}+4abcd+b^{2}c^{2})+6c^{2}d^{2}]S\nonumber\\
+(6a^{2}b^{2}-6c^{2}d^{2})\cos\varphi\nonumber\\
+6(a^{2}bd+ab^{2}c+acd^{2}+bc^{2}d)\sin\varphi\nonumber\\
+6(-a^{2}b^{2}+a^{2}d^{2}+4abcd+b^{2}c^{2}-c^{2}d^{2})\eta=-4r^{2} \tag{36c}%
\end{multline}%
\begin{multline}
4(ab+cd)(b^{2}+d^{2})S+4(b^{3}a-d^{3}c)\cos\varphi+\nonumber\\
2(b^{3}c+3b^{2}ad+3bd^{2}c+ad^{3})\sin\varphi+\nonumber\\
\lbrack-4b^{3}a+6(2d^{2}ab+2cdb^{2})-4d^{3}c]\eta=0 \tag{36d}%
\end{multline}%
\begin{equation}
(b^{2}+d^{2})^{2}S+(b^{4}-d^{4})\cos\varphi+2bd(b^{2}+d^{2})\sin
\varphi+(-b^{4}+6b^{2}d^{2}-d^{4})\eta=2r^{2} \tag{36e}%
\end{equation}
along with eq. (29). Notice that the only difference is that the right hand
sides of eqs.(36a) and (36c) are different from those of eqs. (30a) and (30c).
As we pointed out in eq. (11), there is another symmetry in parameters:
$(\eta,\varphi=\pi/2)\rightarrow(-\eta,\varphi=0)$.\ Hence, above analysis is
also applicable to the case $\eta=-1/4$\ and $\varphi=0$. In fact the
solutions can be reduced to eqs. (13) and (14).

We plot the kink solutions in Fig. 2 for $\eta=0.15$ and $\varphi
=0\symbol{126}1.75\pi$. The solutions at $\eta=0.35$ and $\varphi
=0\symbol{126}0.5\pi$ and $\varphi=0.5\pi\symbol{126}\pi$ are shown in Fig. 3a
and Fig. 3b respectively. The kinks for $\eta<1/4$ (Fig. 2) change smoothly
and retain their shapes as $\varphi$\ moves across $\pi/2$.\ The total change
of $\theta$, $\Delta\theta=\theta(s=\infty)-\theta(s=-\infty)$, is equal to
$2\pi$. This can also be deduced from eq. (28) by tracing the variation of
$\theta$\ with respect to $s.$\ On the other hand, the solutions for
$\eta>1/4$ (Fig. 3a and 3b) develop a second kink as $\varphi$\ approaches
$\pi/2$.\ At $\varphi=\pi/2$,\ the form in eq. (28) is no longer applicable.
Eq. (35) has to be used and it gives a\ large kink and a small kinks which are
the decedents of the connected kinks at $\varphi=\pi/2-\varepsilon$\ where
$\varepsilon$\ is an infinitesimal positive number. Note also that the total
change of $\theta$\ of neither the large kink nor the small kink is equal to
$2\pi$, but rather the sum of them is. This can also be seen from eq. (35).

We here propose a classical explanation of the situation near $\varphi=\pi/2$.
It is related to the minima of the potential. As we have argued in section 2,
$d\theta/ds|_{s=\pm\infty}=0$ and $V(\theta(\pm\infty))=V_{\min}$. The
solutions in eqs. (28) and (35) extend from one minimum to another. The kink
comes from eq. (28) starts from one absolute minimum of the potential, passing
through a major peak and ends at another absolute minimum. In the special case
with the solutions coming from eq. (35), there are a large kink and a small
kink. Both connect two absolute minima but the former passes a major peak of
the potential and the latter passes a minor peak. The large kink extends from
$\theta=2\arctan((b-a)/(d-c))$\ to $\theta=2\arctan((b+a)/(d+c))$\ and the
small kink extend from $\theta=2\arctan((b+a)/(d+c)$\ to $\theta=2\pi
+2\arctan((b-a)/(d-c))$. It is the emergence of additional symmetry in
potential which gives rise to two absolute minima in the range of $2\pi
$\ which in turn, requires two solutions. We will elaborate more on this in
next section.

Finally, we give the form of the bubble solution. Instead of eq. (28), we use
\begin{equation}
f(s)=\frac{a\cosh(rs)+b}{c\cosh(rs)+d} \tag{37}%
\end{equation}
and follow the same procedure, we are able to obtain the bubble solution,
similar to that in eq. (22). Its shape is shown in Fig. 4. by the dashed line
which connects two relative minima of potential. The bubble solution can be
found only when the relative minima exist.

\subsubsection{4. Discussion and conclusion}

We summarize our results on the $\varphi-\eta$\ phase diagram in Fig. 5.\ The
kink solution in eq. (28) can be found in any place except for those vertical
lines. On the vertical lines, the form in eq. (35) prevails. Note that it
corresponds to either $g(s)=\coth(rs)$ or $g(s)=\tanh(rs)$,\ which in turn,
corresponds respectively to large kink or small kink.\ The bubble\ solution
exist in the region above the upper dashed line or below the lower dashed
line, i.e., the regions where relative minima exist.

The limiting case is also interesting. In the limit $\eta\rightarrow0$, the
solutions should be those of an ordinary sine-Gordon equation: $\theta
=4\arctan[\exp(s)]=2\arctan[\csc h(-s)].$\ It is compatible with the form in
eq. (28). In the other limit, $\eta\rightarrow\infty$,\ one can write eq. (1)
as
\begin{equation}
\frac{1}{4\eta}\frac{\partial^{2}\theta^{\prime}}{\partial t^{2}}-\frac
{1}{4\eta}\frac{\partial^{2}\theta}{\partial x^{2}}+\frac{1}{2\eta}\sin
(\theta^{\prime}/2+\varphi)+\sin\theta^{\prime}=0. \tag{38}%
\end{equation}
where $\theta^{\prime}=2\theta$. The third term can be treated as a
perturbation. The zeroth order of the solutions should be those of sine-Gordon
equation with $\sin(2\theta):$\
\begin{equation}
\theta=2\arctan[\exp(s^{\prime})] \tag{39}%
\end{equation}
Here, $s^{\prime}=2\sqrt{\eta}s$, implying $r=2\sqrt{\eta}$\ in eq. (28) as
$\eta\rightarrow\infty$. This is shown clearly by the M\"{o}bius
transformation below.

Consider only the leading order terms of eqs. (30) for large $\eta$%
\begin{equation}
(a^{2}+c^{2})^{2}S+(-a^{4}+6a^{2}c^{2}-c^{4})\eta=0, \tag{40a}%
\end{equation}%
\begin{equation}
4(ab+cd)(a^{2}+c^{2})S+[-4a^{3}b+6(2a^{2}cd+2abc^{2})-4c^{3}d]\eta=0,
\tag{40b}%
\end{equation}%
\begin{equation}
\lbrack6a^{2}b^{2}+2(a^{2}d^{2}+4abcd+b^{2}c^{2})+6c^{2}d^{2}]S+6(-a^{2}%
b^{2}+a^{2}d^{2}+4abcd+b^{2}c^{2}-c^{2}d^{2})\eta=2r^{2} \tag{40c}%
\end{equation}%
\begin{equation}
4(ab+cd)(b^{2}+d^{2})S[-4b^{3}a+6(2d^{2}ab+2cdb^{2})-4d^{3}c]\eta=0 \tag{40d}%
\end{equation}%
\begin{equation}
(b^{2}+d^{2})^{2}S+(-b^{4}+6b^{2}d^{2}-d^{4})\eta=2r^{2}. \tag{40e}%
\end{equation}
It can be shown easily that none of $a$, $b$, $c$ and $d$\ can be 0. Hence we
can set $a=pc$\ and $b=qd$\ where $p$\ and $q$\ are just two ratio parameters.
With the condition eq. (29), it can be shown further that $q=-p=\pm1$\ and
$|cd|=1/\sqrt{2}.$ Now let $S=-\eta+\delta$\ where $\delta=O(\eta^{0})$\ and
substitute it into eq. (40e), one find that $\delta=2r^{2}/d^{4}$.\ This
implies that $b=\pm d=O(\sqrt[4]{\eta})$ and $a=\pm c=O(1/\sqrt[4]{\eta})$. We
thus have shown that for finite $\eta$, solutions of the form of eq. (28)
always exist (except for $(\eta<1/4,\varphi=0)$ and $(\eta>1/4,\varphi=\pi
/2)$) and their parameters $a$, $b$, $c$ and $d$\ vary smoothly with $\eta.$
Hence, the solution for finite $\eta$ has the same form as that for $\eta=0.$

The only places where there are phase transition are the ends of the vertical
lines in Fig. 5, i.e., the points $(\varphi=n\pi,\eta=-1/4)$\ and
$(\varphi=(n+1/2)\pi,\eta=1/4).$\ Here, indeed the form of the solutions
changes from that in eq. (28) into that in eq. (35) when the absolute value of
$\eta$\ increases.\ Classically, this is a second-order phase transition. Its
quantum fluctuation has also been well-studied[12-15].\ 

There is another aspect we would like to investigate and that is varying
$\varphi$ across $\varphi=\pi/2$ for a fixed $\eta.$ We plot in Figure 6a and
6b the energy as a function of $\varphi$\ with $\eta=0.15$\ and $\eta
=0.35$\ respectively. One can immediately notice the behavior of energy near
$\varphi=\pi/2+n\pi$.\ For $\eta=0.15$,\ the slope change is large but still
smooth. For $\eta=0.35$,\ though the energy remains continuous, the slope does
not. This indicates that when $|\eta|<1/4$\ there is smooth crossover. But
when $|\eta|>1/4$\ there is a second-order phase transition from the energy
point of view. The solutions also show different behavior.\ In Fig. 2, the
kinks vary smoothly across the point $\varphi=\pi/2$ at $\eta=0.15$. For the
solutions in Figs. 3 with $\eta=0.35,$ one finds that near $\varphi\approx
\pi/2,$ the shapes of the solutions are different from those at $\varphi
=\pi/2$. The solutions are combination of two kinks though their forms are
still that of eq. (28), i.e. , $g(x)=\sinh(rx)$. As a result the range of
variation of $\theta$ is still $2\pi$. and the topological charge is unity.
When $\varphi=\pi/2$ the form of the kinks is that of eq. (35), i.e.,
$g(x)=\tanh(rx)$ or $g(x)=\coth(rs)$, the large or small kink solutions. It is
the sum of ranges of variation of $\theta$ of the two kinks which is equal to
$2\pi,$\ and thus the deconfinement. It is due to the emergence of the
symmetry that both $\theta$\ and $\pi-\theta$\ are solutions.\ On the other
hand, the solutions at $\varphi=\pi/2\pm\varepsilon$ where $\varepsilon$ is a
infinitesimal number, are very similar. Our numerical results also show that
the coefficients $a$, $b$, $c$, $d$ and $S$\ vary smoothly across the point
$\varphi=\pi/2$, (not including the point $\varphi=\pi/2$.) Thus, the point
$\varphi=\pi/2$\ for $\eta>1/4$ is actually a singular point.

In this work, we used the method of "M\"{o}bius transformation" to solve the
asymmetric DSGE. This method transformed the DSGE into a set of algebraic
equations. Thus we are able to find the forms of the solutions for all the
region on the $\varphi-\eta$ plane. The resulting forms of our solutions can
serve as the basis of various methods, such as form factor perturbation
theory, semi-classical method or a truncated conformal space approach to study
quantum fluctuation.

This work is supported in part by NSC of Taiwan, ROC under the contract number
NSC 95-2112-M-002-048-MY3. One of the authors, C. D. Hu would like to thank
Chern Chyh-Hong for inspiring discussion.

\subsubsection{Appendix: Introduction to M\"{o}bius transformation}

By Jacobi elliptic function theory, one can transform the following equation:%
\begin{equation}
(\frac{df}{ds})^{2}=\varphi(s)=A(f-f_{0})(f-f_{1})(f-f_{2})(f-f_{3}),
\tag{A-1}%
\end{equation}
where $\varphi(s)$ is a polynomial of $s$\ to the three or four power and
$f_{0}$, $f_{1}$, $f_{2}$ and $f_{3}$ are the roots, into the standard form,
i.e., only terms with even powers are present. The \textquotedblright
M\"{o}bius transformation\textquotedblright\ has the form%
\begin{equation}
f=\frac{a\zeta+b}{c\zeta+d}, \tag{A-2}%
\end{equation}
and so does every root%
\begin{equation}
f_{i}=\frac{a\zeta_{i}+b}{c\zeta_{i}+d}. \tag{A-3}%
\end{equation}
If we take special values of $a$, $b$, $c$ and $d$, we can obtain the form%
\begin{equation}
f=\frac{f_{3}(f_{1}-f_{0})\ast\zeta-f_{1}(f_{3}-f_{0})}{(f_{1}-f_{0}%
)\zeta-(f_{3}-f_{0})} \tag{A-4}%
\end{equation}
and eq. (A-1) becomes%
\begin{equation}
(\frac{d\varsigma}{ds})^{2}=B(\varsigma-\beta_{0})(\varsigma-\beta
_{1})(\varsigma-\beta_{2})(\varsigma-\beta_{3}). \tag{A-5}%
\end{equation}
We set%
\begin{equation}
\lambda=\frac{f_{1}-f_{0}}{f_{1}-f_{2}}\frac{f_{3}-f_{2}}{f_{3}-f_{0}}%
=\frac{\beta_{1}-\beta_{0}}{\beta_{1}-\beta_{2}}\frac{\beta_{3}-\beta_{2}%
}{\beta_{3}-\beta_{0}}, \tag{A-6}%
\end{equation}
so that eq. (A-5) becomes%
\begin{equation}
(\frac{d\varsigma}{ds})^{2}=B\varsigma(\varsigma-1)(\lambda\varsigma-1)
\tag{A-7}%
\end{equation}
here $B=A(f_{3}-f_{0})(f_{2}-f_{1}).$ This is the standard form. But we can do
further transform by setting $\varsigma=\xi^{2}$. So eq. (A-7) becomes the
differential equation of Jacobi elliptic function:%
\begin{equation}
(\frac{d\xi}{ds})^{2}=\frac{B}{4}(\xi^{2}-1)(\lambda\xi^{2}-1). \tag{A-8}%
\end{equation}

Instead eqs. (A.4) and (A.8), we can take different transformation by mapping
roots into $1,-1,1/k,$ and $-1/k.$ So%
\begin{equation}
\lambda=\frac{f_{1}-f_{0}}{f_{1}-f_{2}}\frac{f_{3}-f_{2}}{f_{3}-f_{0}}%
=\frac{\frac{1}{k}-1}{\frac{1}{k}+1}\frac{\frac{-1}{k}+1}{\frac{-1}{k}%
-1}=(\frac{1-k}{1+k})^{2} \tag{A-9}%
\end{equation}
and eq. (A-1) becomes%
\begin{equation}
(\frac{d\xi}{ds})^{2}=B^{\prime}(\xi^{2}-1)(k^{2}\xi^{2}-1) \tag{A-10}%
\end{equation}

There are more details to be solved such as how to deal with degenerate roots
and how to further transform $\lambda$ so that it is real and in the range
$(0,1)$ in order that eq. (A-8) and (A-10) are compatible with the Jacobi
elliptic differential equation. But these are beyond the scope of this work.

\paragraph{References:}

[1] S. Coleman, Phys. Rev. \textbf{D 11}, 3424 (1975).

[2] V. E. Korepin and L. D. Faddev, Theor. Mat. Fiz. \textbf{25}, 147 (1975).

[3] A. O. Gogolin, A. A. Nersesyan and A. M. Tsvelik, 'Bosonization and
Strongly Correlated Systems', Cambridge University Press (1998).

[4] T. Giamarchi, 'Quantum Physics in One Dimension', Oxford University Press (2004).

[5]. R. K. Bullough and P. J. Caudrey,'Solitons', Springer-Verlag Berlin
Heidelberg (1980).

[6]. N. Nagaosa, 'Quantum Field Theory in Stongly Correlated Electronic
Systems', Springer-Verlag, Berlin (1999).

[7] R. K. Bullough, J. Mod. Opt., \textbf{47}, 2029 (2000) and erratum,
\textit{idbd}, \textbf{48}, 747 (2001).

[8] C. A. Condat, R. A. Guyerand M. D. Miller, Phys. Rev. \textbf{B 27}, 474 (1983).

[9] D. C. Campbell, J. E. Schonfeld and C. A. Wingate, Physica \textbf{19 D},
165 (1986).

[10] Zuntao Fu, et. al., Z. Naturforsch \textbf{60a}, 301 (2005).

[11] G. Delfino, G. Mussardo and P. Simonetti, Nucl. Phys. \textbf{B 473},.469 (1996).

[12] G. Delfino and G. Mussardo Nucl. Phys. \textbf{B 516},.675 (1998).

[13] Z. Bajnok, L. Palla and G. Tak\'{a}cs, Nucl Phys. \textbf{B 687},.189 (2000).

[14] G. Tak\'{a}cs and F. Wagner, Nucl. Phys. \textbf{B 741},.353 (2006).

[15] G. Mussardo, V. Riva and G. Sotkov, Nucl Phys. \textbf{B 687},.189 (2004).

[16] M. Fabrizo, A. O. Gogolin and A. A. Nersesyan, Nucl. Phys. \textbf{B
580}, 647 (2000)

[17] R. Shindou, J. Phys. Soc. Jpn. \textbf{74}, 1214 (2005).

[18] Nan-Hong Kuo, Sujit Sarkar and C. D. Hu, ,arXiv:0809.2185v1
[cond-mat.str-el] (2008).

\newpage

\paragraph{Figure cations:}

Fig. 1 Potential energy v.s. $\theta$ for $\varphi=0$ and (a) $\eta=-0.35$,
(b) $\eta=0.15$ and (c) $\eta=0.35$.

Fig. 2 Solutions form the form of eq. (28) for $\eta=0.15$ and $\varphi
=0\symbol{126}7\pi/8$.

Fig. 3a Solutions from the form of eq. (28) for $\eta=0.35$ and $\varphi
=0\symbol{126}\pi/2$. At $\varphi=\pi/2$, the solutions have the form of eq.
(35) which gives large and small kinks.

Fig. 3b Solutions from the form of eq. (28) for $\eta=0.35$ and $\varphi
=\pi/2\symbol{126}\pi$.\ At $\varphi=\pi/2$, the solutions have the form of
eq. (35) which gives large and small kinks.

Fig 4. kink solution (in dashed line) and bubble solution (in solid line) for
$\eta=0.35$ and $\varphi=0$. $\theta$\ is in unit of $\pi.$\ 

Fig. 5 Phase diagram on $\varphi-\eta$ plane. See text for explanation.

Fig. 6 The energy of the kink versus $\varphi,$ (a) $\eta=0.15,$ (b)
$\eta=0.35.$

\newpage

Table 1\textbf{\ }List of $a,b,c,d,r,S$ for eq. (28) for.$\varphi=0$

$%
\begin{tabular}
[c]{|c|c|c|c|c|c|c|}\hline
$\eta\diagdown$ & $a$ & $b$ & $c$ & $d$ & $r$ & $S$\\\hline
$-0.25$ & $\times$ & $\times$ & $\times$ & $\times$ & $\times$ &
$0.75$\\\hline
$-0.15$ & $0$ & $0.795271$ & $1.25743$ & $0$ & $\pm0.532456$ & $0.85$\\\hline
$0$ & $0$ & $1$ & $1$ & $0$ & $\pm1$ & $1$\\\hline
$0.15$ & $0$ & $1.12468$ & $0.88914$ & $0$ & $\pm1.26491$ & $1.15$\\\hline
$0.25$ & $0$ & $1.18921$ & $0.840896$ & $0$ & $\pm1.41421$ & $1.25$\\\hline
$0.35$ & $0$ & $1.24467$ & $0.803428$ & $0$ & $\pm1.54919$ & $1.35$\\\hline
$0.45$ & $0$ & $1.29357$ & $0.773055$ & $0$ & $1.67332$ & $1.45$\\\hline
\end{tabular}
$

Table 2 List of $a,b,c,d,r,S$ for eq. (28) for.$\varphi=\pi/4$

$%
\begin{tabular}
[c]{|c|c|c|c|c|c|c|}\hline
$\eta\diagdown$ & $a$ & $b$ & $c$ & $d$ & $r$ & $S$\\\hline
$-0.25$ & $-0.466708$ & $0.836099$ & $0.712245$ & $0.866693$ & $\pm1.26959$ &
$1.10092$\\\hline
$-0.15$ & $-0.460968$ & $0.828478$ & $0.7977$ & $0.735676$ & $\pm1.1254$ &
$1.04093$\\\hline
$0$ & $-0.382683$ & $0.92388$ & $0.92388$ & $0.382683$ & $\pm1$ & $1$\\\hline
$0.15$ & $-0.238106$ & $1.10602$ & $0.890012$ & $0.0656207$ & $\pm1.1254$ &
$1.04093$\\\hline
$0.25$ & $-0.173621$ & $1.20406$ & $0.833646$ & $-0.0216328$ & $\pm1.26959$ &
$1.10092$\\\hline
$0.35$ & $-0.13299$ & $1.2828$ & $0.785775$ & $-0.0600972$ & $\pm1.41323$ &
$1.17461$\\\hline
$0.45$ & $-0.106325$ & $1.34854$ & $0.74763$ & $-0.0772086$ & $\pm1.54803$ &
$1.25625$\\\hline
\end{tabular}
$

Table 3\textbf{\ }List of $a,b,c,d,r,S$ for eq. (28) for.$\varphi=\pi/2$

$%
\begin{tabular}
[c]{|c|c|c|c|c|c|c|}\hline
$\eta\diagdown$ & $a$ & $b$ & $c$ & $d$ & $r$ & $S$\\\hline
$-0.25$ & $-0.594604$ & $0.840896$ & $0.594604$ & $0.840896$ & $\pm1.41421$ &
$1.25$\\\hline
$-0.15$ & $-0.628717$ & $0.795271$ & $0.628717$ & $0.795271$ & $\pm1.26491$ &
$1.15$\\\hline
$0$ & $-0.707107$ & $0.707107$ & $0.707107$ & $0.707107$ & $\pm1$ &
$1$\\\hline
$0.15$ & $-0.88914$ & $0.562341$ & $0.88914$ & $0.562341$ & $\pm0.632456$ &
$0.85$\\\hline
$0.25$ & $\times$ & $\times$ & $\times$ & $\times$ & $\times$ & $\times
$\\\hline
$0.35$ & $\times$ & $\times$ & $\times$ & $\times$ & $\times$ & $\times
$\\\hline
$0.45$ & $\times$ & $\times$ & $\times$ & $\times$ & $\times$ & $\times
$\\\hline
\end{tabular}
$

\newpage%

\begin{figure}
[ptb]
\begin{center}
\includegraphics[
height=4.2151in,
width=5.4129in
]%
{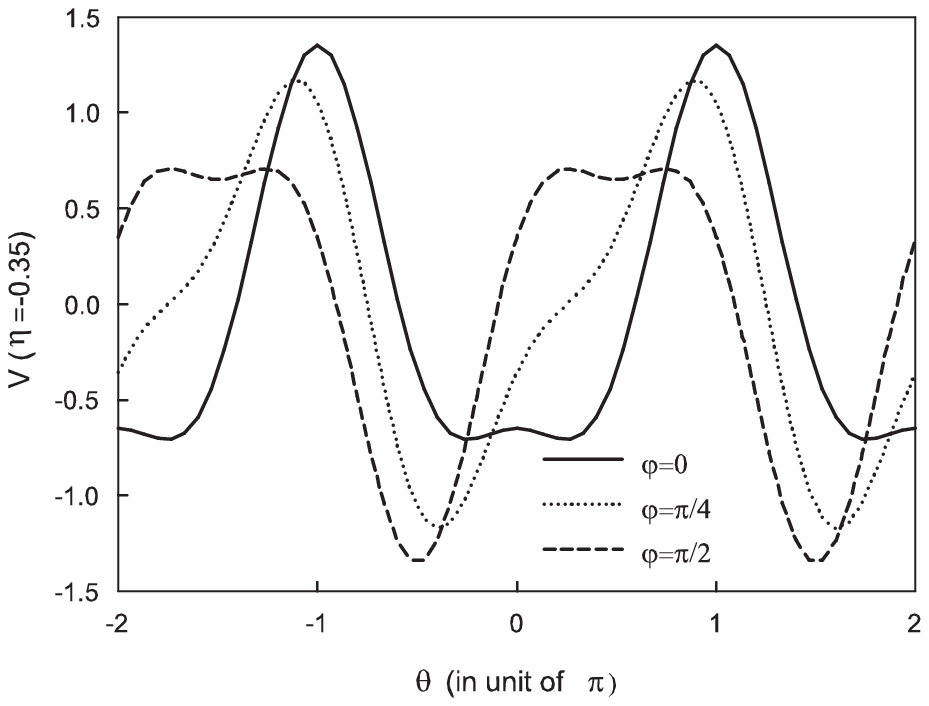}%
\end{center}
\end{figure}
\begin{figure}
[ptb]
\begin{center}
\includegraphics[
height=4.2151in,
width=5.4544in
]%
{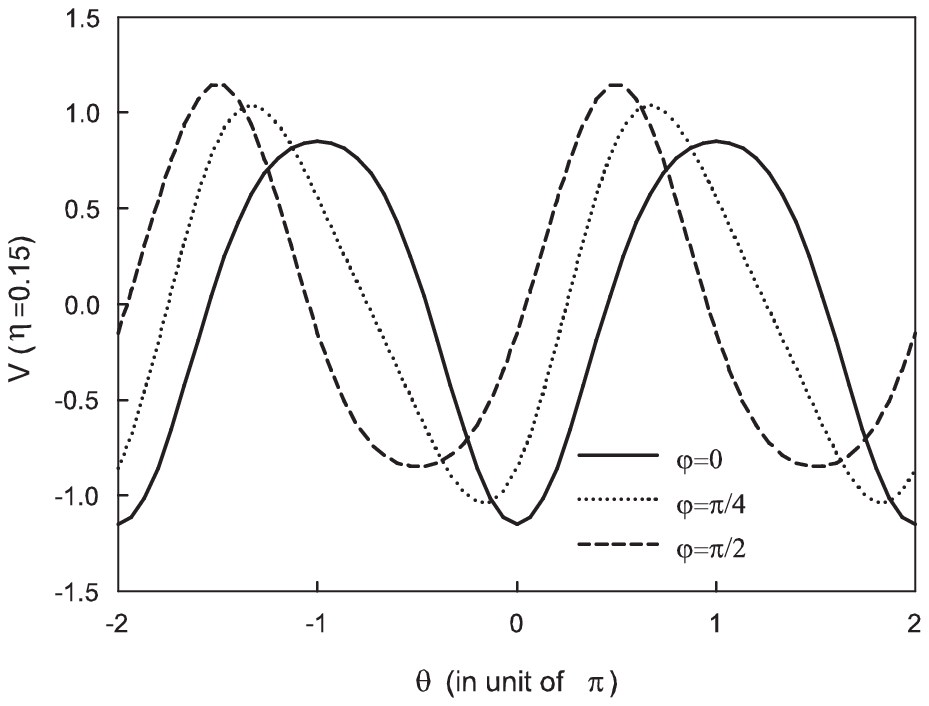}%
\end{center}
\end{figure}
\begin{figure}
[ptb]
\begin{center}
\includegraphics[
height=4.2151in,
width=5.4544in
]%
{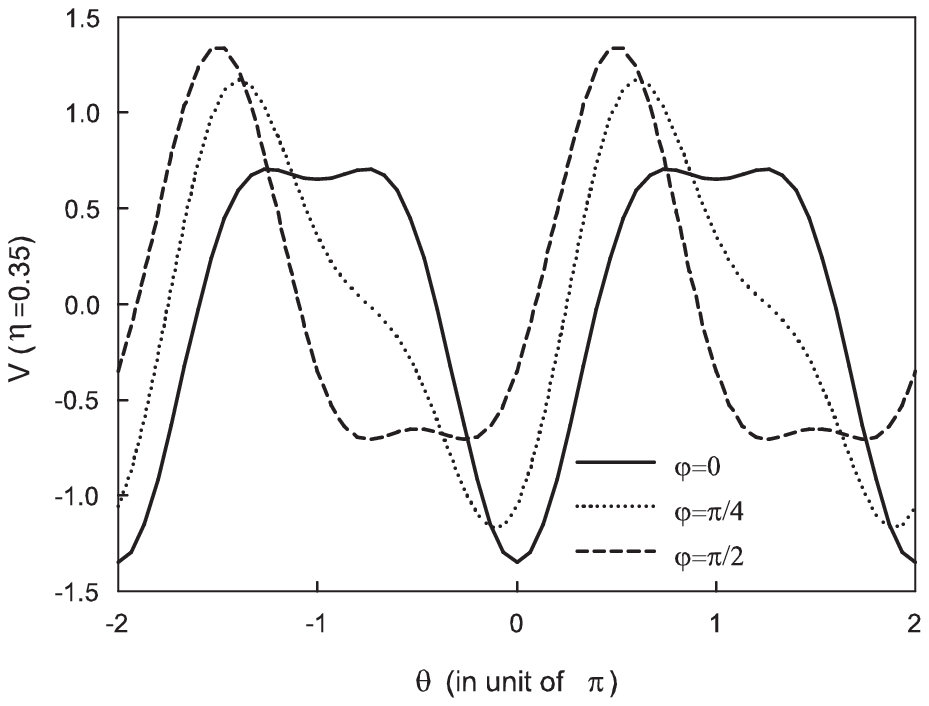}%
\end{center}
\end{figure}
\begin{figure}
[ptb]
\begin{center}
\includegraphics[
height=7.7012in,
width=5.4483in
]%
{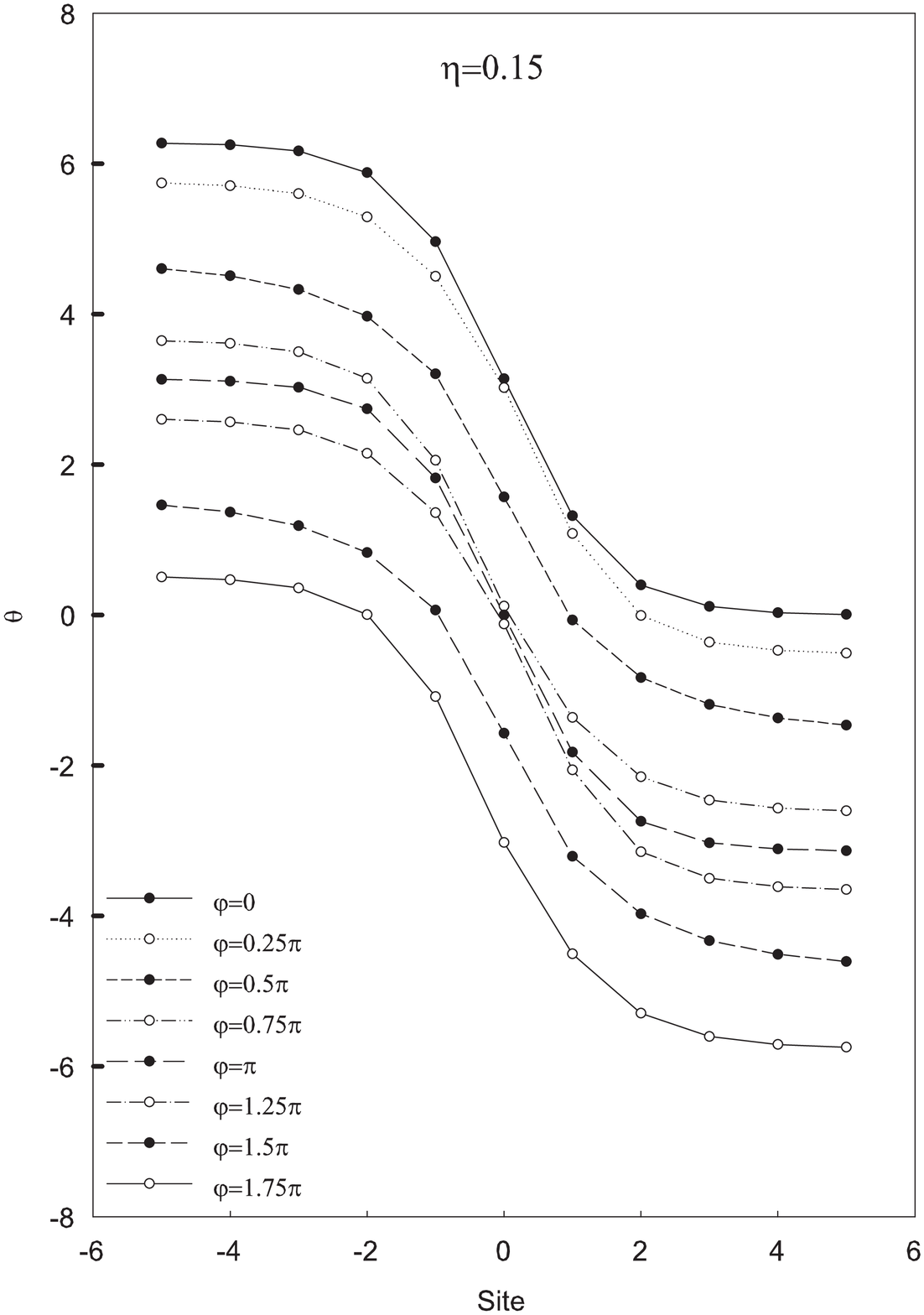}%
\end{center}
\end{figure}
\begin{figure}
[ptb]
\begin{center}
\includegraphics[
height=7.1978in,
width=5.092in
]%
{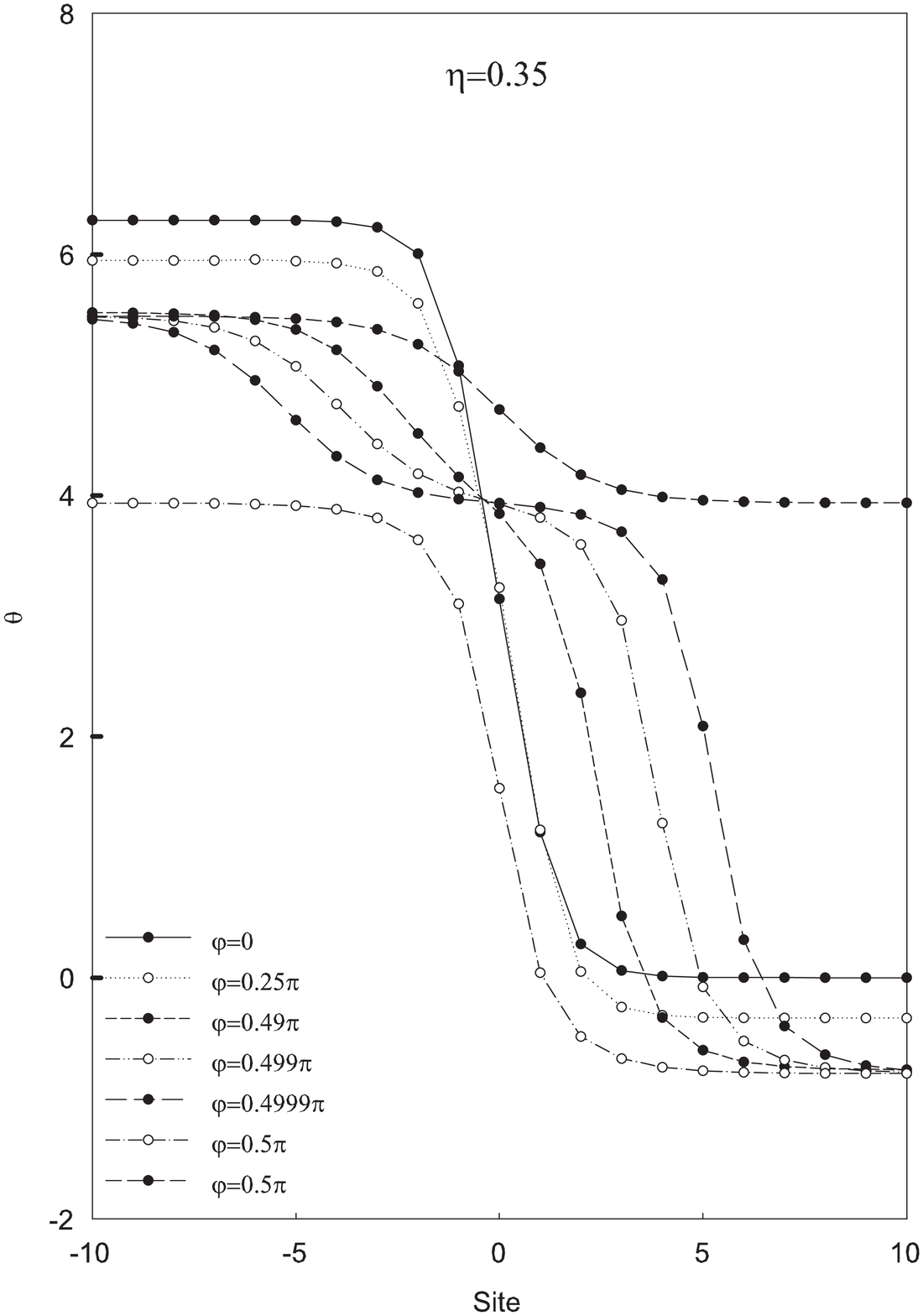}%
\end{center}
\end{figure}
\begin{figure}
[ptb]
\begin{center}
\includegraphics[
height=6.9427in,
width=5.1041in
]%
{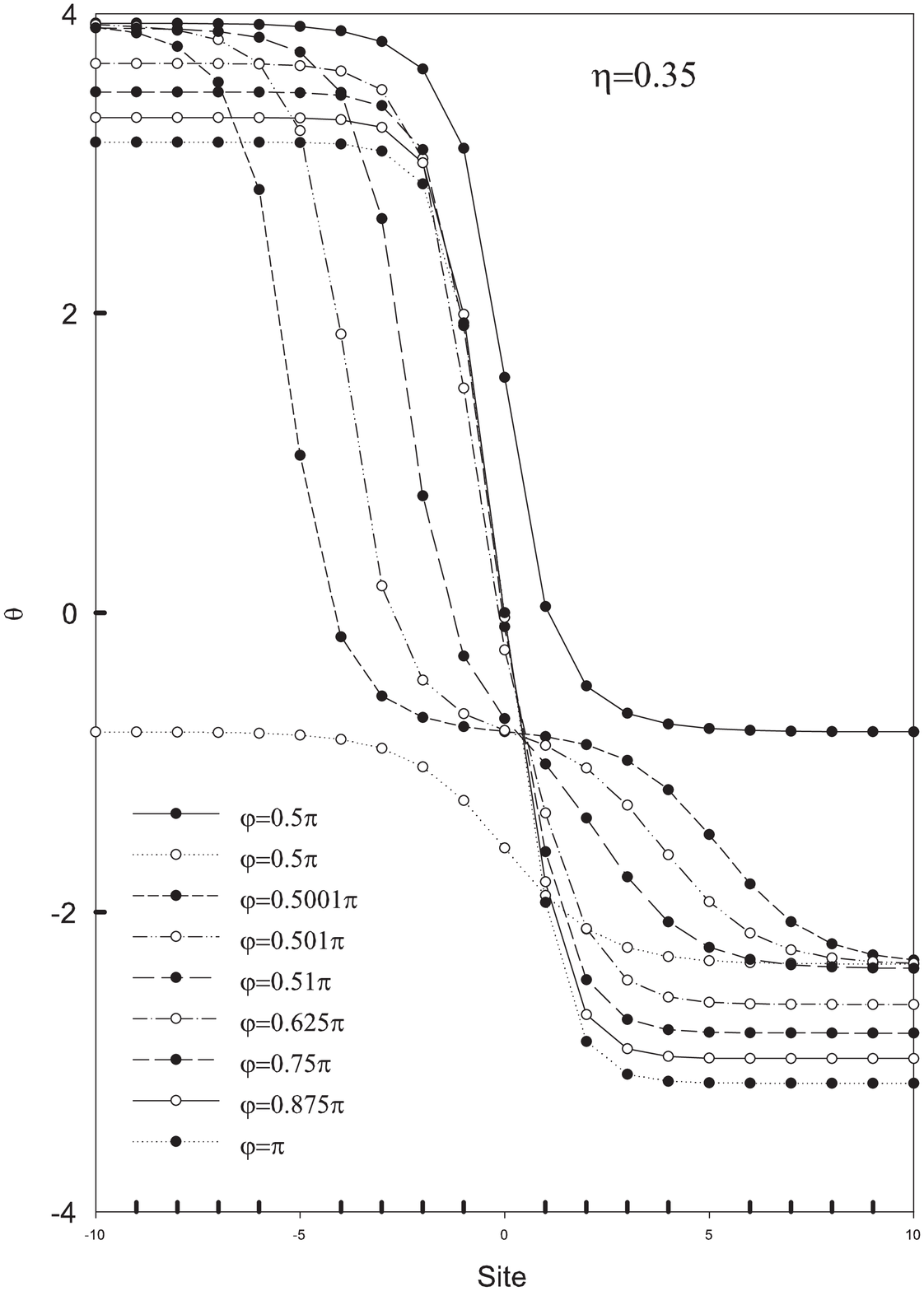}%
\end{center}
\end{figure}
\begin{figure}
[ptb]
\begin{center}
\includegraphics[
height=4.4399in,
width=3.3996in
]%
{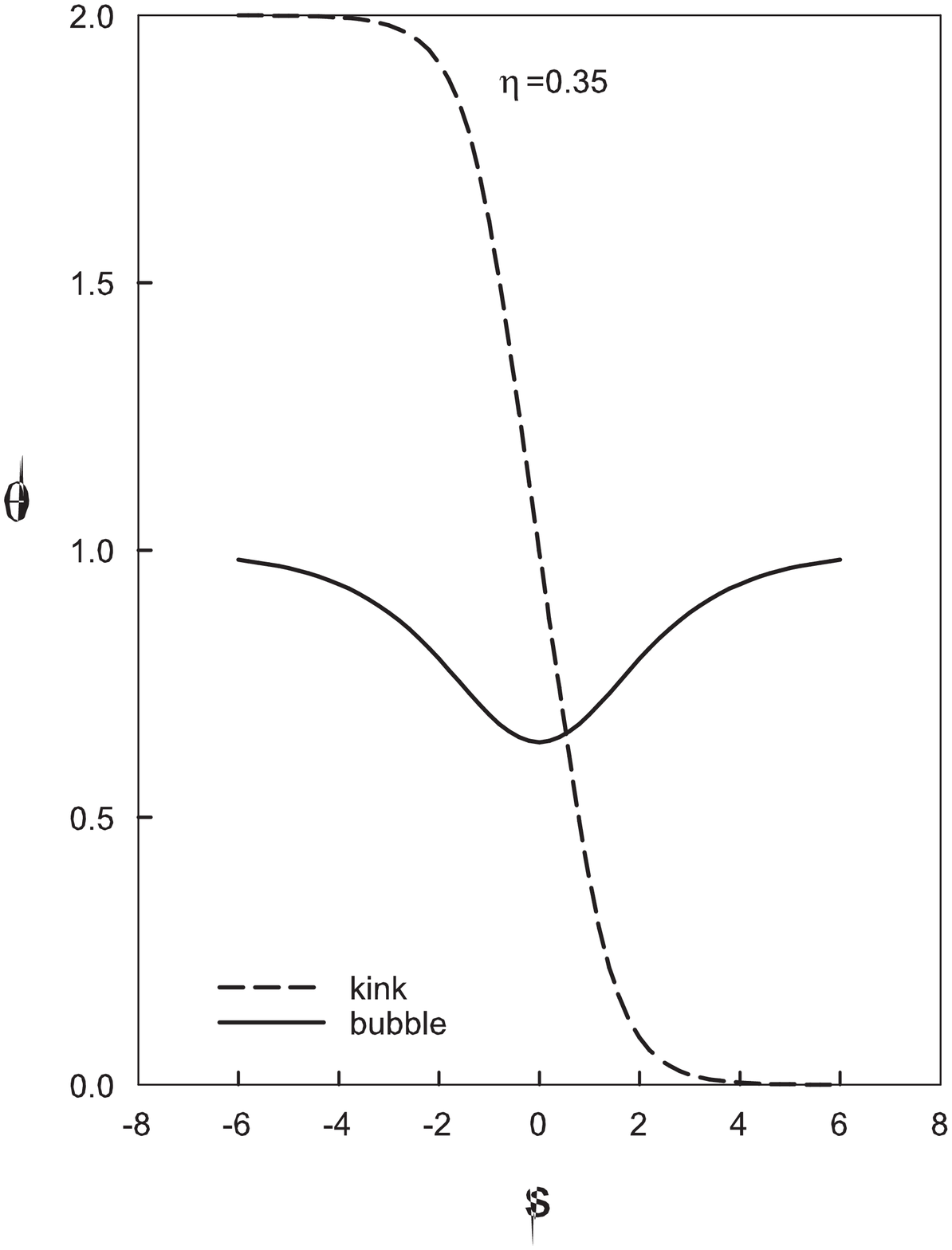}%
\end{center}
\end{figure}
\begin{figure}
[ptb]
\begin{center}
\includegraphics[
height=4.2168in,
width=5.9663in
]%
{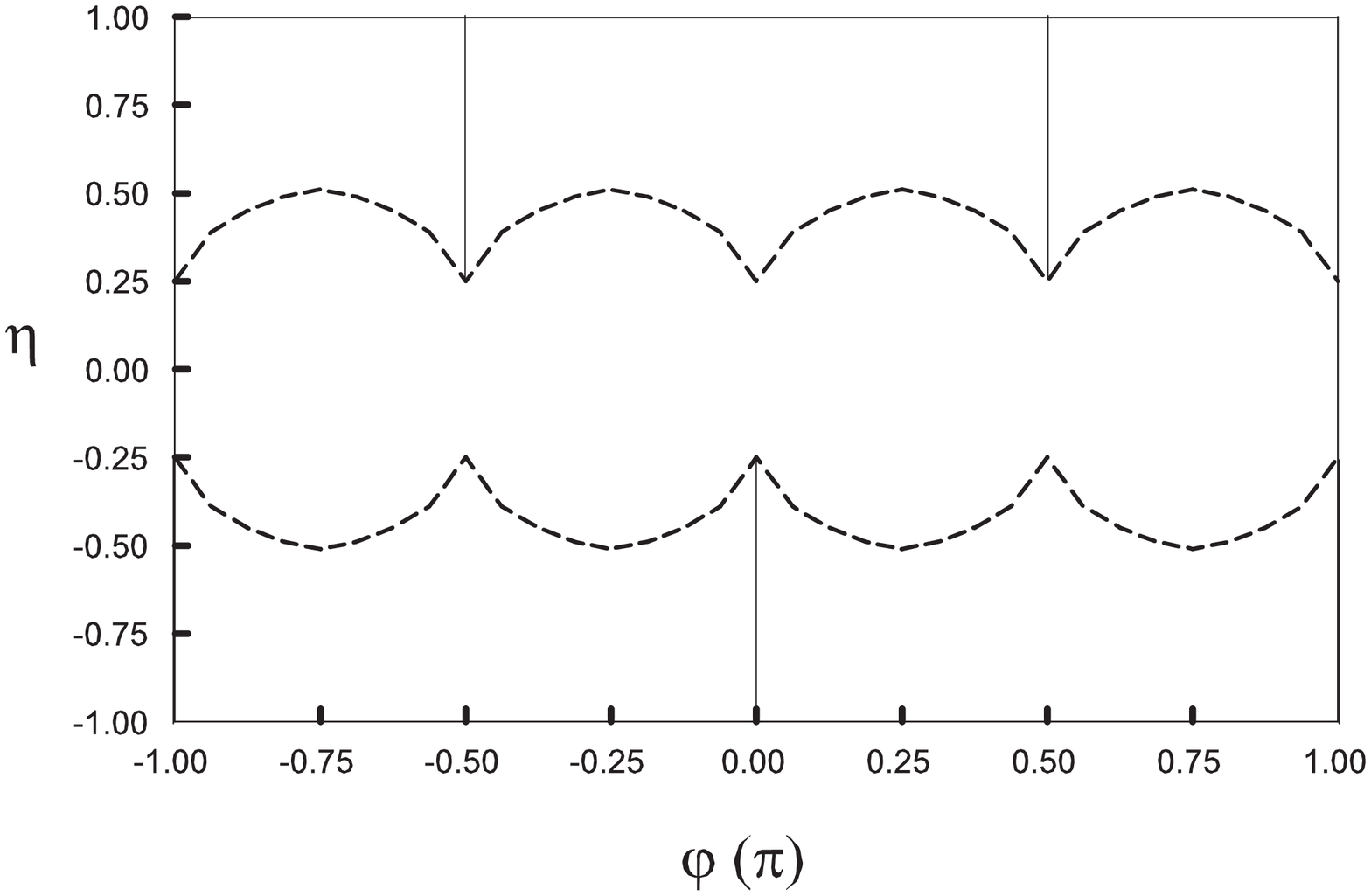}%
\end{center}
\end{figure}
\begin{figure}
[ptb]
\begin{center}
\includegraphics[
height=5.9931in,
width=4.2359in
]%
{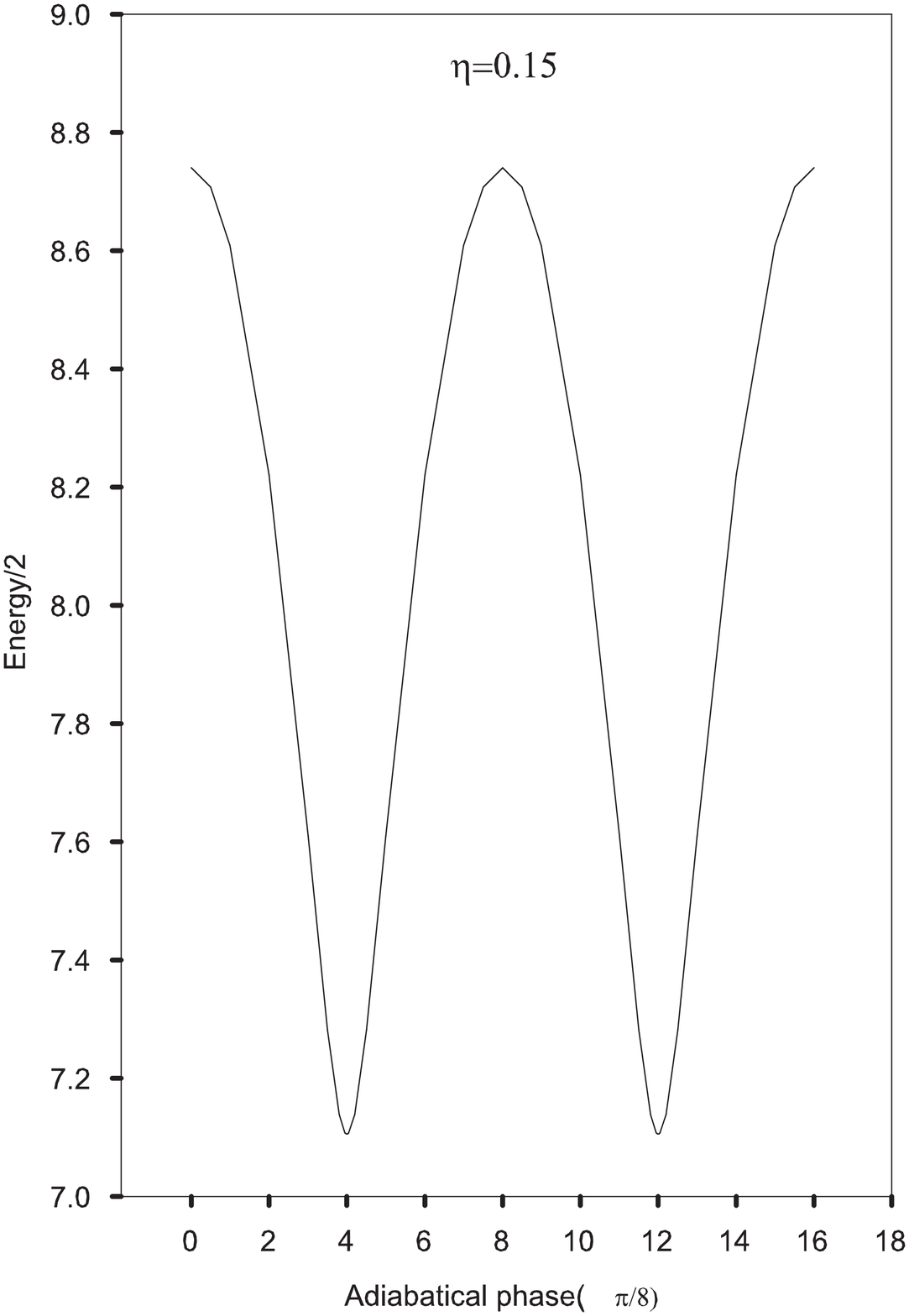}%
\end{center}
\end{figure}
\begin{figure}
[ptb]
\begin{center}
\includegraphics[
height=5.9931in,
width=4.2359in
]%
{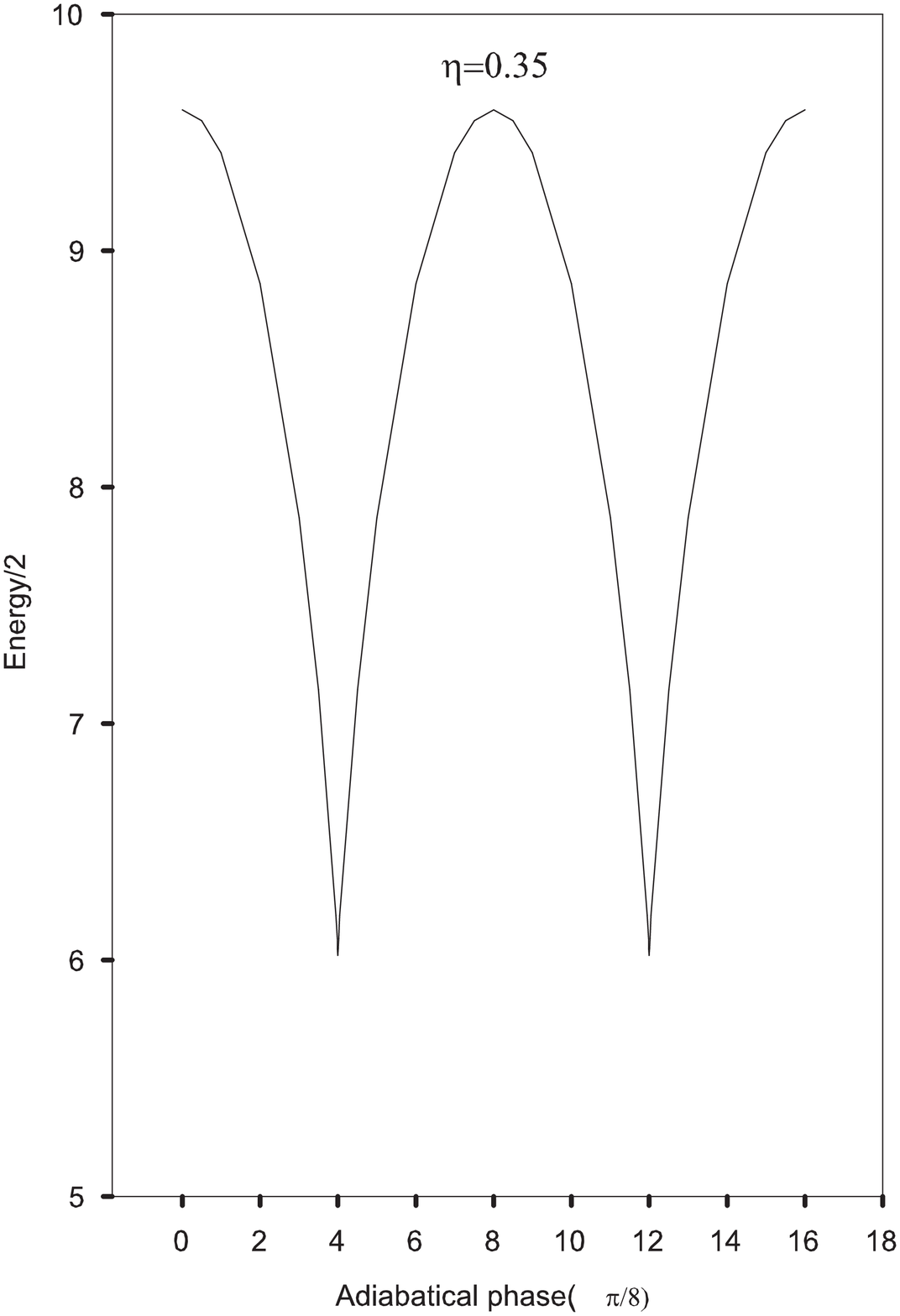}%
\end{center}
\end{figure}

\end{document}